\documentclass[10pt,twocolumn]{article}
\usepackage{iftex}
\ifPDFTeX 
  \usepackage[T1]{fontenc}
  \usepackage[utf8]{inputenc}
  \usepackage{lmodern}
\else 
  \usepackage{fontspec}
\fi
\usepackage[margin=0.75in,columnsep=0.3in]{geometry}
\usepackage{amsmath,amssymb}
\usepackage{graphicx}
\usepackage{booktabs,array}
\usepackage[numbers,sort]{natbib}
\usepackage{microtype}
\usepackage{xurl}
\usepackage[hidelinks]{hyperref}

\title{Fair Feed Ranking for Participatory Budgeting\thanks{Published in
  \emph{Proceedings of the International Conference on Information Technology
  for Social Good (GoodIT '26)}, September 02--04, 2026, Pisa, Italy.
  ACM. \url{https://doi.org/10.1145/3794786.3830735}.
  Licensed under CC BY 4.0.}}
\author{Carina I. Hausladen\\
  University of Konstanz, Konstanz, Germany\\
  \href{mailto:carina.hausladen@uni-konstanz.de}{carina.hausladen@uni-konstanz.de}\\
  ORCID: \href{https://orcid.org/0000-0002-0762-5447}{0000-0002-0762-5447}}
\date{}

\begin{document}
\maketitle

\begin{abstract}
In large-scale participatory budgeting, citizens cannot inspect the full proposal pool, so the order in which proposals are shown becomes a form of agenda-setting power. We argue that fair exposure should therefore be treated as a democratic-design goal. We study Consul Democracy, a widely deployed open-source digital-democracy platform, and show that its proposal feeds are typically ordered by popularity, recency, or comment activity.
Building on this diagnosis, we propose \emph{FairFeed}, a feed-ranking design for PB that uses transparently declared preferences, boosts under-exposed proposals, and admits a rate-limited reject channel for crowd-sourced vetting. 
We evaluate the design in a simulation anchored in Munich's 2025 PB process and compare it with random, newest, and most-commented feeds.
In this simulation, FairFeed broadens proposal discovery, distributes visibility more evenly across the eligible pool, increases cross-cutting support, and improves resistance to manipulation relative to comment-based ranking. We conclude by outlining the human-subjects evaluation needed to test whether onboarding can recover voter preferences accurately enough for deployment in practice.
\end{abstract}


\section{Introduction}

Participatory budgeting (PB) allows residents to decide directly how public funds are spent. Digital platforms have expanded PB from small in-person assemblies to large-scale processes involving thousands of participants and proposals.
A defining feature of PB is its openness. Unlike formal elections, participation is often not restricted by citizenship, voter registration, or even adult age~\cite{chicago_49th_ward_pb,cambridge_pb_2026}. Lowering these barriers is one of PB's central democratic achievements.
Yet this success creates a new challenge. As participation expands, so does the number of proposals competing for public attention: recent PB cycles draw proposal pools in the thousands, with Valencia alone reaching 2,776 submissions in 2025 (Table~\ref{tab:scale}).

\paragraph{Scale creates two linked problems: attention and vetting.}

The first problem is attention. Citizens cannot realistically inspect thousands of proposals before voting. This limitation is particularly important online, where users rarely proceed beyond the first page of results~\cite{urman2023search}.
As a consequence, the set of proposals a voter encounters is shaped by two agenda-setting forces: the platform's ordering rule and the voter's entry point into the process, such as a proposal recommended by friends or campaign networks. These forces operate on two margins. On the \emph{extensive margin}, does a voter encounter proposals beyond the one that initially brought them to the platform? On the \emph{intensive margin}, is attention concentrated on already-visible proposals, or do less prominent but potentially valuable proposals receive meaningful exposure?
A democratic process is not judged only by how many people can participate, but by what kind of participation it enables. If PB platforms fail on either margin, they deny citizens a meaningful chance to encounter alternatives and form considered judgments~\cite{cohen2009deliberation,fishkin2011people}. Fair exposure is therefore best understood as a floor on visibility: without it, a proposal's failure may reflect rejection, but it may equally reflect lack of exposure.

The second problem is vetting. As participation expands, submissions become more numerous and more heterogeneous in quality. Every proposal must be assessed for feasibility, legality, cost, and duplication. Centralized review is not only costly, but it also concentrates agenda-setting power in the hands of administrators. Delegating evaluation back to citizens avoids this concentration of power, yet immediately reintroduces the attention problem.

Existing PB processes illustrate both dimensions of the problem (Table~\ref{tab:scale}): administrators have to screen large submission pools; yet the proposal lists citizens must navigate still run into the hundreds.
The problem is even aggravated by how cities try to help citizens navigate through these large lists:
Consul Democracy (originally CONSUL, developed by the City of Madrid in 2015) is one of the most widely adopted open-source digital democracy platforms. It is used by more than 200 public institutions in over 35 countries and serves more than 100 million people.\footnote{\url{https://consuldemocracy.org/}} 
The Consul Democracy web interface lists proposals in a grid that can be sorted; out of the box the project feed offers two orderings, \emph{most-voted} and \emph{random}~\cite{consul_source}. 
A deployment can widen this menu. However, the orderings cities offer are dominated by popularity, comment-count, and recency signals, all of which amplify already-visible proposals (Table~\ref{tab:feeds}).

\begin{table}[t]
  \centering
  \footnotesize
  \setlength{\tabcolsep}{4pt}
  \renewcommand{\arraystretch}{1.0}
  \caption{PB at scale.}
  \label{tab:scale}
  \begin{tabular}{@{}lrrr@{}}
    \toprule
     & & \multicolumn{1}{c}{Admins} & \multicolumn{1}{c}{Citizens} \\
    \cmidrule(lr){3-3}\cmidrule(lr){4-4}
    Process (cycle) & Submitted & Reviewed & Ballot \\
    \midrule
    Barcelona '20--23 & ---     & 823     & 204 \\
    Paris '25         & 2{,}079 & ---     & 261 \\
    Valencia '25--26  & 2{,}776 & 1{,}441 & 458 \\
    München '25       & 1{,}043 & 1{,}043 & 459\,$\to$\,20 \\
    \bottomrule
  \end{tabular}
  \begin{flushleft}
    \footnotesize
    Dashes mark stages not separately reported.
    Sources:~\cite{barcelona2021_pb,paris2025laureats,paris2025budget_participatif_vote,valencia_analysis2526,mb2025_unser}.
  \end{flushleft}
\end{table}

\begin{table}[t]
  \centering
  \footnotesize
  \setlength{\tabcolsep}{3pt}
  \renewcommand{\arraystretch}{1.0}
  \caption{Feed orderings offered.}
  \label{tab:feeds}
  \begin{tabular}{@{}>{\raggedright\arraybackslash}p{0.40\columnwidth}cccc@{}}
    \toprule
     & \multicolumn{1}{c}{Neutral} & \multicolumn{1}{c}{Recency} & \multicolumn{2}{c}{Engagement} \\
    \cmidrule(lr){2-2}\cmidrule(lr){3-3}\cmidrule(lr){4-5}
    Deployment & Random & Newest & Comments & Votes \\
    \midrule
    Consul (stock) & $\star$ & & & $\bullet$ \\
    \midrule
    \raggedright Cancún, Glasgow, Inverclyde, Renfrewshire, N.~Lanarkshire, Fryslân, Aarhus & $\star$ & & & \\
    Valencia & $\star$ & $\bullet$ & & \\
    \raggedright Marchamalo, Valladolid, Getafe, São Paulo & $\star$ & & & $\bullet$ \\
    München '25, Garching, Jena & $\star$ & $\bullet$ & $\bullet$ & \\
    Gelsenkirchen & $\star$ & $\bullet$ & & $\bullet$ \\
    München '26, Unterschleißheim & $\star$ & $\bullet$ & $\bullet$ & $\bullet$ \\
    \bottomrule
  \end{tabular}
  \begin{flushleft}
    \footnotesize
    $\star$~default, $\bullet$~offered; Votes~= most-voted (incl.\ live
    ranking). Stock Consul from~\cite{consul_source}; the rest audited by the
    authors from the live deployments, with per-deployment URLs in the
    reproduction package.
  \end{flushleft}
\end{table}

\paragraph{Learning from recommender systems.}
The attention problem is not unique to PB. E-commerce platforms, streaming services, and social media all face the challenge of helping users navigate spaces that are far too large to inspect manually. Recommender systems were developed precisely to allocate scarce attention.
These systems operate on both margins identified above. On the \emph{extensive margin}, recommendations help users discover items they would otherwise never encounter. Recommendations drive about a third of Amazon purchases~\cite{mackenzie2013amazon}, while roughly 70\% of YouTube watch time and 80\% of Netflix viewing originate from recommended content~\cite{solsman2018youtube,gomezuribe2015netflix}. A mechanism built to sell more products may also serve a democratic function: citizens typically enter PB through a proposal for their own neighborhood, and widening what they encounter can broaden their horizon toward other communities' needs.

On the \emph{intensive margin}, recommender systems determine which items occupy a limited attention window. This is crucial because online attention is highly concentrated at the top of rankings~\cite{carnovalini2025popularity}. Feedback therefore becomes biased toward whatever is already visible, and rankings based on prior engagement tend to generate ``rich-get-richer'' dynamics~\cite{chen2019offpolicy}. For PB, this implies that a desirable ranking system should actively correct for unequal exposure. More specifically, this problem has been formalized in ranking research as fairness of exposure, equity of attention, and expected exposure, which allocate visibility across items in proportion to merit over one or many rankings rather than letting early popularity determine who is seen~\cite{singh2018fairness,biega2018equity,diaz2020evaluating}. However, current PB platforms actually aggravate the issue with their implemented rankings (Table~\ref{tab:feeds}).

\paragraph{Learning from online communities.}
Large-scale PB also requires mechanisms for identifying weak, infeasible, duplicate, or low-quality proposals. 
Online communities provide useful precedents. Stack Overflow and Reddit pair approval with downvotes so that the community itself can demote low-quality, off-topic, or duplicate content. However, downvoting is typically used sparingly~\cite{StackOverflow2025WhatOverflow}.
A related idea already exists within PB. The D21--Janeček method supplements positive votes with a limited number of negative votes, while requiring each voter to cast more positive than negative votes~\cite{janecek2021d21}. 
Paris goes further: since 2021 its budget vote uses majority judgment, where each voter rates every proposal on a four-point scale from strong endorsement down to an explicit ``not convinced;'' however, there is no cap on that negative mention~\cite{paris2025budget_participatif_vote}.
For PB processes involving thousands of submissions, optional negative feedback can act as crowd-sourced vetting.

\paragraph{Contributions.}
This paper contributes both a diagnosis and a design. Conceptually, it argues that in large-scale PB, feed ranking is a form of agenda-setting power and that fair exposure should be treated as a democratic-design goal.
Empirically, it audits Consul deployments and field data from MünchenBudget to show that currently deployed orderings rely heavily on popularity, recency, and especially comment signals that are sparse and vulnerable to manipulation. 
Building on this diagnosis, it proposes \emph{FairFeed} and evaluates its ranking logic in a simulated PB environment. The design uses declared preferences, exposure correction, and a rate-limited reject channel to broaden discovery, equalize visibility, and increase resilience against attempts to manipulate rankings, for example through coordinated commenting.

\section{Methods}

\subsection{FairFeed}
\begin{figure}
    \centering
    \includegraphics[width=\linewidth]{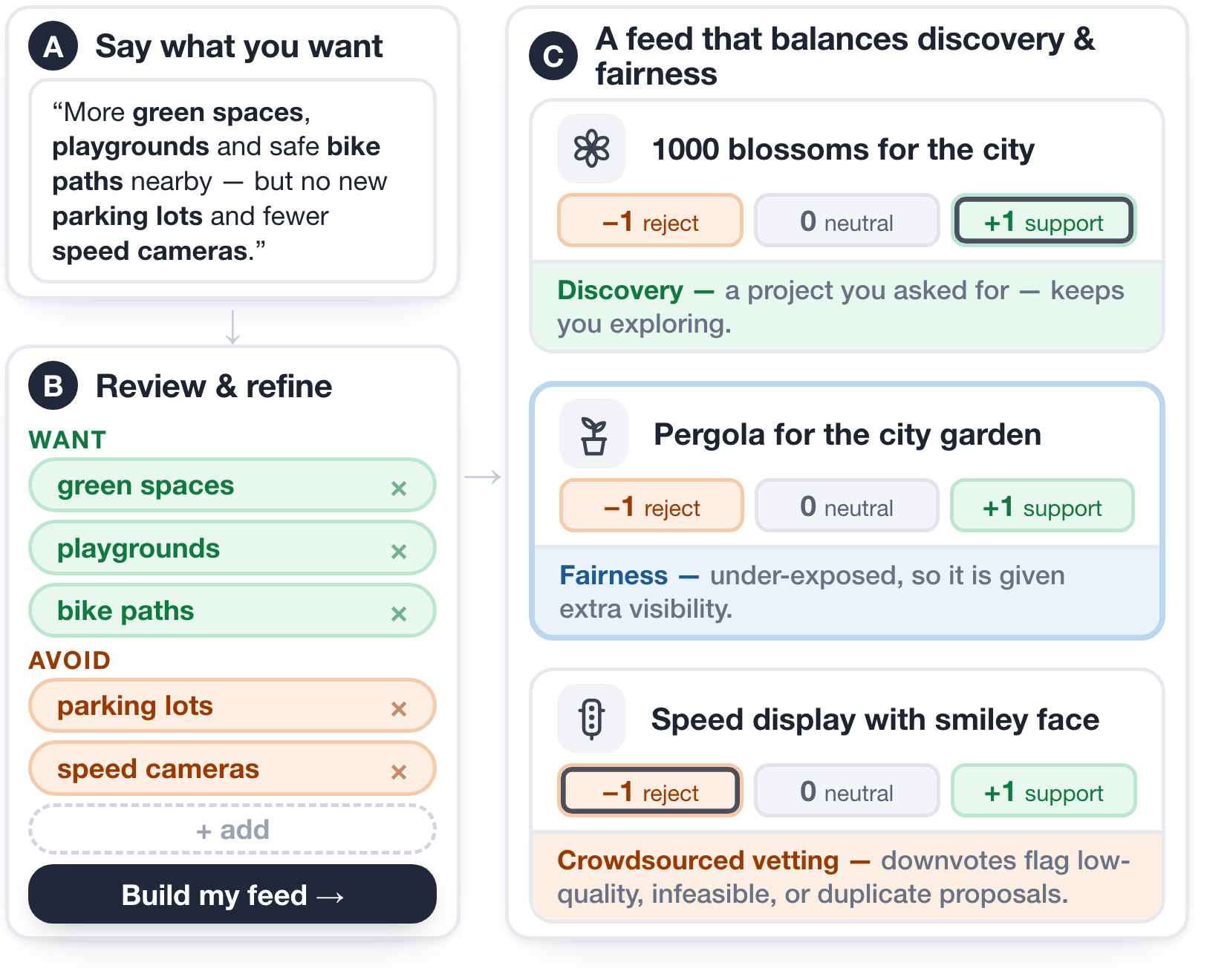}
\caption{FairFeed onboarding prototype: the citizen describes interests in natural language, which an LLM maps to want/avoid signals \emph{(A)}; reviews and refines them \emph{(B)}; and then receives a personalised feed that boosts underexposed ones (blue frame), and admits downvotes for crowdsourced vetting \emph{(C)}.}
\label{fig:onboarding}
\end{figure}

The observations outlined above motivate the design of \emph{FairFeed}, a recommender system for large-scale PB (Figure~\ref{fig:onboarding}).
FairFeed addresses the attention problem through \emph{exposure correction}: it deliberately lifts under-exposed proposals so that attention is not monopolized by projects that benefit from early mobilization or ranking advantages.
It addresses the vetting problem through \emph{negative feedback}. Citizens can not only approve or remain neutral toward a proposal but also reject it; this allows them to register concerns about proposal quality, feasibility, or duplication and thereby contribute to crowd-sourced quality control. Importantly, this negative channel is guarded: under the D21--Janeček rule, a citizen can cast a reject only after supporting two other proposals~\cite{janecek2021d21}.
Finally, because recommendation is inherently agenda-setting, the ranking process must be \emph{transparent}, contestable, and responsive to citizens' stated preferences. In FairFeed, proposals are ranked only on declared signals \cite{malki2025bonsai}.
FairFeed has three parts: The onboarding interface (Figure~\ref{fig:onboarding}) is a working prototype; the ranking rule (Eq.~\ref{eq:fairfeed}) is the design we evaluate; and the voter model (Eqs.~\ref{eq:support}--\ref{eq:reject}) is a simulation standing in for real users, proxying elicited preferences with broad topic categories. What we test here is the ranking rule; the interface and preference elicitation await the human-subjects study (Section~\ref{sec:conclusion}).

\subsection{Data}
\label{sec:data}
Public access to election-related data is a recognised democratic-transparency norm~\cite{carolan2017open}.
In that spirit, we draw our data from \emph{MünchenBudget}, one of the largest municipal PB schemes on Consul in Germany, with a \texteuro{}1{,}000{,}000 budget in 2025 and 2026 each. 
From the completed 2025 round (1,043 proposals) and the ongoing 2026 round (1,059 proposals), we download, for each proposal, its identifier and pseudonymised author, and for each comment, its pseudonymised author.

Of the 1{,}043 proposals submitted in 2025, 459 passed admissibility and were displayed in the Consul interface for citizen likes; the 20 most-liked proposals advanced to the final vote (Table~\ref{tab:scale}). Thus, ballot inclusion depended directly on proposal ranking in the interface.
Across both rounds, comments in \emph{MünchenBudget} represent a sparse engagement signal: 69\% of 2025 proposals received no comments. Nevertheless, comment activity increased substantially, with self-comments rising from 13\% in 2025 to 38\% in 2026. This increase is largely driven by a single account, which produced around 40\% of all 2026 comments, primarily on its own proposals and those of one other account.
The following simulation is calibrated to these empirical patterns.


\subsection{Modeling PB}
\label{sec:model}
We develop a simulation calibrated to the \emph{MünchenBudget} data. Behavioral assumptions follow prior literature: position decay~\cite{craswell2008click,meredith2013ballot} and the divisiveness discount~\cite{hausladen2024legitimacy}. The remaining parameters (the support threshold $\tau$, the fit slope $\beta$, and the browse-length distribution) are transparent normalizations or dials held \emph{fixed across feeds}, so no free parameter can drive the cross-feed differences we report.
Generating such a model allows us to quantify both the negative effects of existing feed-sorting algorithms and the potential benefits of FairFeed.

\paragraph{The proposals.}
There are $M=459$ proposals (the proposals listed for citizens in 2025) and $N=7{,}134$ citizens, the participant count MünchenBudget reported for its 2025 round.
Each proposal $j$ is assigned a latent quality $q_j\in[0,1]$. To anchor the simulation to Munich's 2025 outcome, the $14$ proposals with recorded outcomes are calibrated directly: the $10$ funded proposals receive quality scores obtained by linearly mapping their observed vote counts to the interval $[0.70,1]$, while the four vetoed proposals are assigned $q_j=0.55$. The remaining $445$ proposals are synthetic, with quality drawn from $q\sim\mathrm{Beta}(2,5)$. 
Each proposal also receives a reach parameter $\kappa_j$, representing the audience its author can mobilize, drawn from $\mathrm{Pareto}(1.5)+1$, a topic category $c_j$, and a divisiveness score $\xi_j$ drawn from $\mathrm{Beta}(1.5,4)$. For synthetic proposals we assume quality and reach independent ($q\perp\kappa$), so boosting low-reach items is costless in surfaced quality: the unseen are, on average, no worse than the seen. This is consequential: a positive correlation could let the boost surface weaker items. As a check we re-pair reach and quality among the synthetic proposals to impose $\rho\in\{0.3,0.6\}$ (both marginals held fixed): the least-seen quartile does weaken (mean $q$ from $0.31$ at $\rho{=}0$ to $0.28$ at $\rho{=}0.6$), yet top-20 quality holds or rises ($+22\%$ at $\rho{=}0.3$) because support still gates on $\tau_i$ (Eq.~\ref{eq:support}). The independence assumption is thus conservative for the shortlist claim; a fuller $(q,\kappa,\xi)$ sweep remains future work (Section~\ref{sec:conclusion}).
Comments are not a fixed attribute but accrue as citizens browse: each card a voter views draws a comment with probability $0.04\,(1+3\xi_j)$ (rank-decayed for browsed cards), so divisive proposals attract proportionally more discussion. The most-commented feed ranks on this running tally.

\paragraph{The voters.}
Citizens arrive sequentially. Each voter $i$ enters the platform through a single proposal $s_i$, the \emph{seed} that motivated the visit. Seeds are drawn with probability proportional to proposal reach, $\Pr[s_i=j]=\kappa_j/\sum_{\ell}\kappa_\ell$, so proposals with greater visibility attract more visitors. The seed distribution is held fixed across feed designs. We assume feed ranking affects which proposals citizens discover, and may affect turnout~\cite{peixoto2017gotv}, but not their underlying preferences, so any cross-feed difference reads as a ranking effect. This assumption is load-bearing and only partly supported (our citation concerns turnout, not preference formation); we flag preference drift as a threat to validity (Section~\ref{sec:conclusion}).

After arriving, voter $i$ browses $V_i$ proposal cards returned by the feed. We interpret $V_i$ as the number of cards scanned while scrolling, not the number of proposals opened and read in full. We model $V_i$ as geometric with mean 10.

For each proposal encountered, the voter casts support if its quality exceeds a personal threshold $\tau_i$, centred at $0.5$, the midpoint of the quality scale.
Supports cast beyond the seed are the \emph{cross-cutting votes} that form our principal outcome.
Relevance is defined at the seed's broad topic category: a non-seed proposal is on-topic ($m_{ij}=1$) when it shares that category and off-topic ($m_{ij}=0.3$) otherwise. A feed that improves discovery exposes voters to more such proposals and generates more cross-cutting votes.
We treat topical interests as exogenous: the seed proposal's category proxies the interests a real onboarding process would elicit.


\paragraph{The voter's decision.}
The seed is supported automatically. 
For each other proposal $j$ shown at rank $r$, the voter supports it with probability
\begin{equation}
P^{\mathrm{sup}}_{ij}(r)=\underbrace{\sigma\!\big(\beta\,(q_j m_{ij}-\tau_i)\big)}_{\text{fit}}\;\underbrace{(1-\lambda\xi_j)}_{\text{divisiveness}}\;\underbrace{\phi(r)}_{\text{position}}.
\label{eq:support}
\end{equation}
The three factors each lie in $[0,1]$. 
\emph{Fit} asks whether quality clears the voter's bar $\tau_i$. Relevance-weighted quality $q_j m_{ij}$ counts in full on-topic ($m_{ij}=1$) and drops to under a third otherwise ($m_{ij}=0.3$), so an off-topic proposal must be roughly three times as good. 
The logistic $\sigma(z)=(1+e^{-z})^{-1}$ then turns the gap $q_j m_{ij}-\tau_i$, scaled by the slope $\beta$, into a support probability.
\emph{Divisiveness} discounts contested proposals by at most half ($\lambda=0.5$): we model a proposal that splits voters (high $\xi_j$) as drawing opposition from one faction, so it accumulates less net support than an equally good consensus proposal. Divisiveness is a distinct axis from quality~\cite{hausladen2024legitimacy}.
\emph{Position} captures that voters read the top and skim the rest, $\phi(r)=1/(1+kr)$ with $k=0.15$, so a proposal at rank 10 carries weight $0.40$, under half the top of the feed. This decay is calibrated between web-search position bias~\cite{craswell2008click} and ballot-order primacy~\cite{meredith2013ballot}. 
The dials $(\beta,\lambda,k)=(5,0.5,0.15)$ are fixed across feeds, so any difference comes from the feed, not the voters.

\paragraph{How long they browse.} 
The budget $V_i$ above is fixed, which isolates the ranking effect: the same attention, only redistributed. But attention is not fixed. 
Information-foraging accounts hold that people linger while returns stay high and leave when they drop~\cite{pirolli1999foraging}.
We therefore also let browsing length respond to the feed: after each item the citizen reads on with a probability that rises with the recent relevance of what she has seen and falls with fatigue, $\Pr[\text{continue past rank } r]=\sigma\!\big(\beta_0+\beta_R\,\mathrm{eng}_r-\beta_F\,r\big)$~\refstepcounter{equation}(\theequation)\label{eq:continue}, where $\mathrm{eng}_r$ is an exponential moving average of recent fit ($\alpha=0.6$) and $(\beta_0,\beta_R,\beta_F)=(0.5,4.0,0.05)$. 
Under this rule a relevance-matched feed holds attention longer, so \emph{how far citizens explore} becomes an outcome in its own right; as the coefficients are from the literature, not fit to MünchenBudget, we read it for direction, not magnitude.
Which rule applies depends on the outcome: the discovery and exposure results (Figure~\ref{fig:results}a,b) let browsing respond to the feed, while the support, cross-cutting-vote, and quality results (Figure~\ref{fig:results}c,d) hold $V_i$ fixed, so any feed difference there is ranking alone, not attention.

\paragraph{Support, coverage, and the shortlist.}
FairFeed's card offers a three-point rating: reject ($-1$), neutral ($0$), support ($+1$). This gives an explicit negative and a neutral midpoint, with a single positive grade. Each support adds one to a proposal's accumulated support $x_j$.
Exposure is a separate, binary signal: every browsed card registers exactly one of the three ratings, and any of them (support, neutral, or reject) makes the proposal \emph{seen} and adds one to its exposure $e_j$.
The neutral point is what keeps exposure correction fair.
Exposure gives \emph{coverage}: the share of proposals clearing a visibility floor ($e_j\geq\bar{e}/2$).
The top-scoring proposals form the \emph{shortlist}: the at-most-20 reaching the decisive vote, whose mean quality $q$ measures how well the ballot tracks merit. 

\paragraph{The reject channel.}
As we argue above, FairFeed offers a negative vote.
We model a gated reject: the negative option unlocks only after a voter has rated a minimum number of items. 
A browsed proposal $j$ at rank $r$ that the voter does not support is then rejected with probability $P^{\mathrm{rej}}_{ij}(r)=\bar{r}\;\xi_j\,(1-q_j m_{ij})\,\phi(r)$~\refstepcounter{equation}(\theequation)\label{eq:reject}. Rejections fall where you would expect: on divisive items ($\xi_j$ high) that fit the voter poorly ($q_j m_{ij}$ low) and sit high enough to be seen. 
$\bar{r}\in[0,1]$ is the \emph{cast-rate}, how often citizens actually reject.

\paragraph{The feeds.}
A feed rule orders the proposals a voter sees. We compare the three baseline orderings most relevant to Consul deployments and Munich in particular. \emph{Random}, one of Consul's stock orderings, reshuffles each visit. \emph{Newest}, which Munich offers, sorts by submission date. \emph{Most-commented}, another ordering Munich offers, is also a popularity sort: it ranks on the running comment count, so comments lift a proposal, its higher position draws more views, and more views generate more comments, a self-feeding popularity loop that is the engine of all three failures (concentrated attention, low-quality proposals on the ballot, and an attack surface for manufactured comments).

\emph{FairFeed} breaks the loop by reading no popularity signal. It scores each proposal on declared quantities only,
\begin{equation}
s^{\mathrm{FF}}_{ij}=\underbrace{m_{ij}}_{\text{topical match}}+\underbrace{b\,\mathbf{1}[e_j<\bar{e}/2]}_{\text{under-exposure boost}}-\underbrace{w_r\,\tfrac{R_j}{\max_\ell R_\ell+1}}_{\text{reject penalty}},
\label{eq:fairfeed}
\end{equation}
and ranks proposals by descending $s^{\mathrm{FF}}_{ij}$. The \emph{topical match} $m_{ij}\in\{1,0.3\}$ carries full weight when $c_j=c_{s_i}$ and a third otherwise. The \emph{under-exposure boost} ($b=0.5$) lifts any proposal seen less than half as often as the average ($e_j<\bar{e}/2$), a relative floor that scales with turnout. The \emph{reject penalty} ($w_r=1.5$) subtracts accumulated rejects $R_j$, normalised by the current maximum. FairFeed never reads engagement volume, which is the channel the popularity loop travels through.

\section{Results}
\label{sec:simulation}

\begin{figure*}[t]
  \centering
  \includegraphics[width=\textwidth, trim={0 8 0 0}, clip]{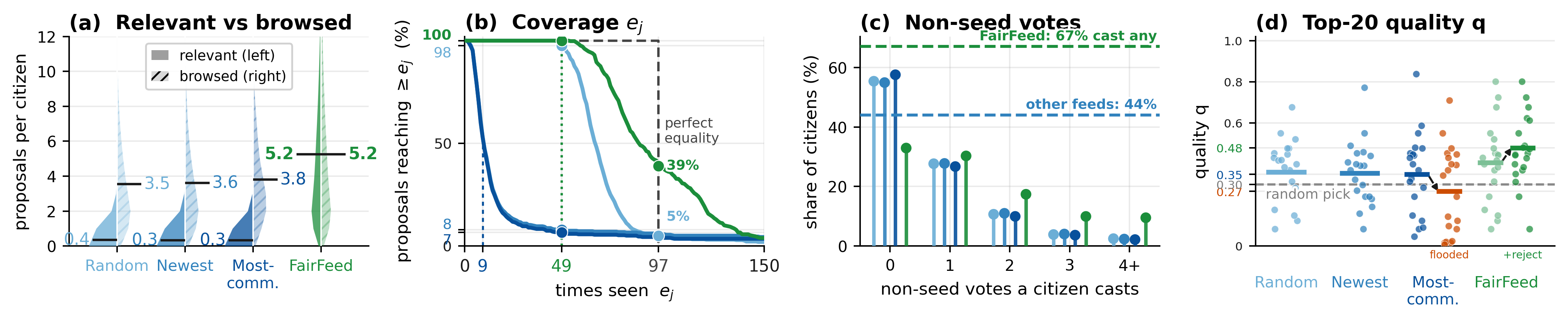}
  \caption{FairFeed (green) browses more and more-relevant projects \emph{(a)}, spreads exposure so most projects clear the visibility floor \emph{(b)}, roughly doubles cross-cutting votes \emph{(c)}, and via its gated reject lifts top-20 quality \emph{(d)}.}
  \label{fig:results}
\end{figure*}

\paragraph{Citizens browse more on topic proposals under FairFeed.}
Assuming citizens continue browsing only while the proposals they encounter remain relevant (Eq.~\ref{eq:continue}), current feed designs sustain at most about $3.8$ viewed proposals per citizen. Yet only $\approx 0.3$ of these are on-topic ($m_{ij}=1$, i.e.\ $c_j=c_{s_i}$). FairFeed, by contrast, increases exposure to approximately $5.2$ proposals per citizen, while maintaining near-perfect topical relevance throughout (Figure~\ref{fig:results}a).

\paragraph{Popularity sorts starve most proposals of exposure.}
Spreading FairFeed's total exposure (each citizen's automatically-supported seed plus the $\approx\!5.2$ cards she browses) evenly across \emph{MünchenBudget}'s $7{,}134$ citizens and $459$ proposals would show each proposal about $97$ times; the visibility floor sits at half that, $\approx 49$ views, the same threshold $\bar{e}/2$ at which FairFeed's under-exposure boost activates (Eq.~\ref{eq:fairfeed}), so panel (b) reports exactly how many proposals the boost rule treats as under-exposed.
Newest and most-commented feeds fall far short of this ideal: they pile views on a few proposals and leave the median proposal seen only nine times.
FairFeed and random instead spread attention more evenly, so almost every proposal clears this floor: all of them under FairFeed and $98\%$ under random, against just $7$--$8\%$ under the popularity sorts (Figure~\ref{fig:results}b).
Clearing the floor only means a proposal escaped obscurity, not that it got its fair share: at the perfect-equality benchmark ($\approx 97$ views) $39\%$ of proposals reach it under FairFeed but only $5\%$ under random, whose relevance-blind feed cuts browsing short (Figure~\ref{fig:results}b).

\paragraph{Boosting underseen proposals costs no relevance.}
Exposure correction is usually a trade: more visibility for under-seen items, less relevance for the user. In our simulation it is not.
For the following results, rather than assume browsing lengthens with relevance, the browse budget is held fixed. This is less realistic, but it isolates cleanly that any difference in support traces to \emph{which} proposals a feed ranks high, not how long it holds the reader.
With $459$ proposals across ten topics a citizen has about $46$ on-topic proposals available but browses only $\approx10$, and an on-topic proposal outranks an off-topic one whether boosted or not. 
The under-exposure boost therefore reorders \emph{within} the citizen's topic. 
Across the run $99.5\%$ of browsed proposals are on-topic (and $97\%$ of the boosted ones); the rare off-topic browse falls to the heaviest browsers, who exhaust their topic ($V>46$, under $1\%$ of citizens). This holds because our topics are broad: ranking on declared topical match could reinforce a bubble within a narrow enough interest, so whether it survives once real voters describe finer interests is the empirical crux for the study (Section~\ref{sec:conclusion}).

\paragraph{On-topic exposure converts to support.}
We model each citizen as arriving for a single project, a neighbour's proposal encoded as the seed $s_i$ (Section~\ref{sec:model}), which they back before leaving. A feed's democratic value is whether it moves them past that entry point to consider, compare, and support other proposals. We measure this as the \emph{cross-cutting} votes a citizen casts beyond her seed.
FairFeed gets far more citizens to branch out: each casts $1.4$ cross-cutting votes on average, against about $0.7$ under random, newest and most-commented and $67\%$ cast at least one, against about $44\%$ (Figure~\ref{fig:results}c).

Why does FairFeed draw out these extra votes, and are they real preferences or clicks on whatever sits on top? To separate earned support from placement we read the support model's first factor alone, $\mathrm{fit}=\sigma(\beta(q_j m_{ij}-\tau_i))$, which omits the position weight $\phi(r)$: a high-fit vote is one a citizen would cast at any rank. Under our parameters clearing $\mathrm{fit}\ge0.5$ requires an on-topic proposal that also passes $q_j\ge0.5$, so the test shows whether a vote is placement-robust, though it cannot fully separate genuine preference from topical match, the very signal FairFeed ranks on. By this measure FairFeed's extra votes are placement-robust: $35\%$ clear the bar against $12\%$ under most-commented, whose top slots hold popular but off-topic proposals backed on position, not relevance.

\paragraph{FairFeed's gated reject cleans the ballot.}
MünchenBudget is approve-only; FairFeed adds a rate-limited reject whose negative option unlocks only after two approvals (the D21--Janeček rule~\cite{janecek2021d21}) and ranks the ballot on net support (supports minus rejects). 
How much it cleans depends on how often citizens reject; this in turn depends on the interface: an optional button draws a small minority (on Stack Overflow, downvotes are only about a tenth of votes~\cite{StackOverflow2025WhatOverflow}), a forced choice far more (on Tinder, where every profile forces a swipe, women swipe right on under $30\%$ and reject the rest~\cite{swipestats2025}). 

At an optional-reject-button rate ($\bar{r}\approx0.10$) top-20 quality rises $+18\%$ over FairFeed's own approve-only ballot, from $0.41$ to $0.48$ (Figure~\ref{fig:results}d). The reject also makes the shortlist less divisive (lower $\xi_j$). A divisive proposal is one voters split on: many back it, many would strike it. This is a separate axis from whether it is good. 
Average divisiveness on the shortlist falls $-21\%$.
Because the reject rule (Eq.~\ref{eq:reject}) is keyed to high $\xi_j$, this drop is partly mechanical; we read it as a sanity check that the channel reaches the proposals Munich's admins actually struck. The clearest case is the two would-be-vetoed proposals we calibrated as decent but highly divisive ($\xi_j>0.6$): an approve-only feed cannot remove them, the reject gives opponents a channel, and one drops off. This cuts both ways: divisive proposals are often where genuine community trade-offs are being contested, and deliberative theory would treat that contestation as material for debate~\cite{cohen2009deliberation}. Whether the gated reject removes weak items or suppresses legitimately contested ones is a line only a human-subjects study can draw; the D21 2:1 gate and the support-outweighs-opposition rule are meant to keep the channel a rationed minority veto, not a majority silencer~\cite{janecek2021d21}.

Additionally, simulating the full range of reject-rates reveals a knee at $\bar{r}\approx0.30$: quality lifts to $+32\%$ and divisiveness to $-35\%$, after which both flatten. A cast-share near $0.30$ is therefore the sweet spot that user-interface designers can aim for.

\paragraph{Negative feedback inverts the incentive to game.}
We report a manipulation analysis to harden the design against gaming, not to enable it.
Ranking on a net signal invites an obvious worry: downvote rivals to bury them.
The D21--Janeček rule counteracts this: a reject unlocks only after two approvals~\cite{janecek2021d21}.
To measure that cost, we simulate an attack that pushes a low-quality proposal onto the ballot. Without a reject channel, the attacker only approves its own proposal. That takes a median of $36$ verified identities. With a weaponised reject strategy, it also downvotes rivals. That lowers the cost to $25$ identities ($0.35\%$): easier, but still not cheap.
This is the contrast with the comment channel. One account can post without limit, as the actor behind $\approx40\%$ of Munich's 2026 comments did. In the worst case, manufactured comments on weak proposals drop most-commented's top-20 quality from $0.35$ to $0.27$ (Figure~\ref{fig:results}d).
Rejects scale only with distinct verified identities. Even pure vandalism displaces half the shortlist only at about $75$ identities ($1.1\%$ of voters); below about $20$, only a few slots move.

\section{Conclusion and Outlook}
\label{sec:conclusion}

This paper treats feed ranking in large-scale PB as a democratic design question: once proposal pools reach the hundreds or thousands, PB faces both an \emph{attention} problem (citizens cannot inspect the full pool, so ranking decides what is seen) and a \emph{vetting} problem (weak, duplicate, or divisive proposals are hard to filter without re-centralising agenda-setting power).

FairFeed is a ranking design aimed at both problems at once. It ranks on declared rather than engagement-based signals, using citizens' stated interests to surface relevant proposals, boosting under-exposed proposals so that visibility is distributed more fairly across the pool, and admitting a rationed reject channel so that citizens can contribute to crowd-sourced vetting.

Evaluated in a model anchored on MünchenBudget field data, FairFeed browses more proposals and more relevant ones, spreads exposure across the eligible pool, roughly doubles cross-cutting support, and via its gated reject improves shortlist quality, with the D21--Janeček 2:1 gate making that negative channel materially harder to game than Munich's current most-commented sort.

This matters for practice: trust in civic procedures is hard to build and easy to lose, so making the hidden consequences of ranking rules visible before deployment is a useful first step.

Because the current simulation proxies elicited preferences with broad topic categories, the next step is to deploy the existing prototype in a human-subjects evaluation. FairFeed's main open empirical questions are whether onboarding recovers voters' real interests with enough accuracy, whether transparent preference signals increase trust and perceived agency, and whether exposure correction remains cheap once real interests are narrower and more fragmented than our stylised topic categories. Deployment also raises institutional questions, especially around contestability, auditing, and identity verification for the reject channel. The broader issue is not only whether FairFeed improves ranking metrics, but whether civic platforms can make agenda-setting power more visible and more evenly distributed in practice.

\paragraph{Code and data availability.} The simulation code and public calibration data are released at \url{https://github.com/carinahausladen/fairfeed-sim-for-PB} (code MIT, data CC-BY-4.0).

\newcommand{\weblink}[2]{\url{#1}}
\bibliographystyle{plainnat}
\bibliography{references}

\end{document}